\documentclass[12pt]{article} 
\usepackage{amssymb}
\usepackage{graphicx}
\begin{document}  
\begin{center} 
{\large PICOSEC: Charged particle Timing to 24 picosecond Precision with MicroPattern Gas Detectors }


				Sebastian White
\footnote{
representing the PICOSEC Collaboration:
European Organization for Nuclear Research (CERN), Geneva 1211, Switzerland;    
University of Virginia, Charlottesville, Virginia 22903, USA;
CEA, IRFU, Centre d'Etudes Nucléaires de Saclay, Gif-sur-Yvette 91191, France ;
 Univ. of Santiago de Campostela, Santiago de Campostela, Spain;
 University of Science and Technology of China, Hefei ,China;
 Aristotle University of Thessaloniki, Greece;
 NCSR Demokritos, Athens, Greece}
\end{center}

\begin{abstract}
	The prospect of pileup induced backgrounds at the High Luminosity LHC (HL-LHC) has
stimulated intense interest in technology for charged particle timing at high rates. 
In contrast to the role of timing for particle identification, which has driven incremental improvements in timing,
the LHC timing challenge dictates a specific level of timing performance- roughly 20-30 picoseconds.
Since the elapsed time for an LHC bunch crossing (with standard design book parameters) has an rms spread of 170 picoseconds,
the $\sim50-100$ picosecond resolution now commonly achieved in TOF systems would be insufficient to resolve multiple ``in-time" pileup.
Here we present a MicroMegas based structure which achieves the required time precision (ie 24 picoseconds for 150 GeV $\mu$'s) and could potentially offer
an inexpensive solution covering large areas with $\sim 1$ cm$^2$ pixel size. We present here a proof-of-principle which motivates 
further work in our group toward realizing a practical design capable of long-term survival in a high rate experiment.
	\end{abstract}

\section{Introduction}

	In 2015 our collaboration presented\cite{Thomas} initial studies of single and multiple photoelectron timing using the structure shown schematically in figure \ref{fig-cartoon}.
In that paper, which used Laser measurements (similar to those presented below) and a simpler entrance window (1mm thick Quartz and a 10 nm thick Al photocathode), it was shown
that single photoelectron time jitter of $100-200$ picoseconds could be achieved. Furthermore the time jitter was shown to scale as $~1/\sqrt{N_{photoelectrons}}$ by varying the laser intensity (and hence the mean number
of photoelectrons). 

	It was therefore plausible 
that charged particle timing to the precision needed for HL-LHC could be achieved using Cerenkov photons produced in MgF2 (having good transparency at $\lambda\sim 200$ nm) and a semitransparent CsI photocathode. 

	It should be noted that this approach imposes a minimum pixel size since the Cerenkov cone radius is roughly equal to the radiator thickness ($\sim3$mm MgF2 in our case).  This granularity is appropriate for the ``Hermetic Timing" approach adopted by the CMS experiment\cite{seb}- ie acceptance for a large fraction of the tracks
produced in an event in order to associate track vertices with a ``time of interaction". This tagging provides a powerful time reference (t$_0$) in order to properly associate other physics objects with time tags in the event.

	A popular buzz-word these days is ``4D-tracking", which clearly contrasts to this approach. So long as the cost-per-channel of timing is significantly higher than that of tracking pixels, timing will have its own granularity dictated by occupancy.
	
	In 2016 we tested several devices in 3 testbeam campaigns at the CERN SPS H4 beamline.
Referring to the initial chamber described in Ref.\cite{Thomas}, we performed primarily measurements with - Ne($80\%)/CF_4(10\%)/C_2H_6(10\%)$ at 1050 mbarr- called ``COMPASS" mixture below.

	The laser measurements we report here focussed on a detailed mapping of the single photoelectron time response vs. the Electric field configuration and particularly
the single photoelectron signal mean time of arrival and time spread. As will be described in a forthcoming paper, the charged particle time resolution can be derived from these results- where the fundamental limit arises from longitudinal diffusion (and therefore time spread) over a distance roughly equal to that from photoemission to the first preamplification step.
It is perhaps remarkable that we have now moved beyond the past paradigms where the discussion was about electronic resolution (ie SNR and t$_{Rise}$) to a focus on detector physics.
	
\section{Results}

	For the results presented below the E-field in both the drift/preamplification region and Mmegas/amplification region (see Fig. 1 left) resulted in roughly equal gain of $\sim 10^2-10^3$ in both stages. Because the Mmegas signal is input to a BW=2GHz current amplifier the signal due to electron and ion collection in the last stage are clearly resolved (Fig. 1 right). We perform typically $20-30\%$ constant fraction timing on the digitized leading elelectron component of the signal. The PIOCSEC time is then relative to a t$_0$ reference (Si Diode in the case of laser data and a Micro Channel Plate based beam counter with $\sim4$ picosecond measured time jitter for the SPS $\mu$-beam measurements).
\begin{figure}
\centering
\centerline{\includegraphics[width=15cm, height=5cm]{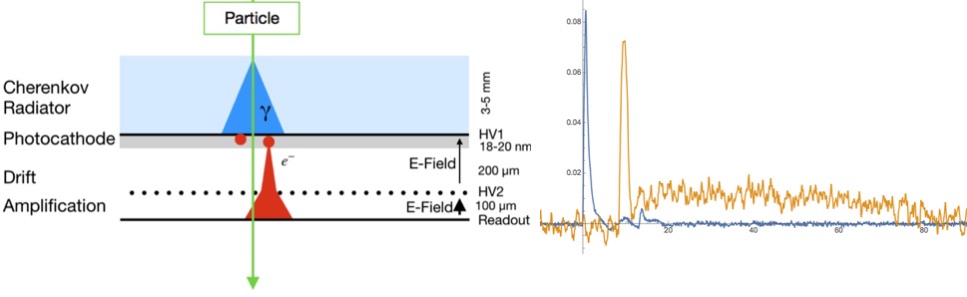}}
\caption{Basic principle of the PICOSEC detector used in these measurements(left). The oscilloscope trace (right-nsec timebase) is taken from a 120 fsec laser pulse where a large fraction of the light pulse is injected into 
a Silicon Diode time reference (blue waveform) and a small fraction (yielding a single photoelectron in this case) impinges on the PICOSEC photocathode.  Note the rapid leading component of the PICOSEC signal which is due to the electron component of the Micromegas signal- on which our Constant Fraction timing is performed. }
\label{fig-cartoon}       
\end{figure}

	For the laser measurements, which were performed at the Saclay IRAMIS SLIC facility, the laser intensity was tuned to give a low ($\sim0.5\%$) occupancy to prevent contamination from multiple photoelectron signals. The results presented in Fig.2 (left) summarize the averages over the full signal spectrum (so-called ``Polya" statistical distribution) due to single photoelectrons for a given field configuration. Nevertheless a detailed study of the mean time of arrival and jitter vs. amplitude was performed and, as reported in a forthcoming paper, elucidates the detector physics which controls timing performance.
	
	PICOSEC is able to give single photoelectron time resolution of $\sim75$ picoseconds, which is perhaps interesting in itself. There are several low rate experiments which could benefit from a thin, large area photodetector with such time resolution (for example to distinguish Cerenkov and scintillation light components- ie CHESS and Theia).
	
	Of course an ongoing emphasis in PICOSEC is to find rugged, high quantum efficiency (or, possibly, secondary emitters) which can turn this good photoelectron timing result into a roughly 3 times better charged particle time jitter.

	As shown in the Figure (2-right) we have achieved this already with an 18nm thick transparent CsI photocathode deposited on a 5.5 nm Cr substrate. In addition to this particular photocathode, beam tests were performed with
various metal and CVD diamond photocathodes as well as variants (in thickness) of the CsI photocathodes. The main lifetime limitation of the CsI photocathode is, as is well known, related to ``ion back-flow" and mitigating these effects, including possibly with protective layers is an area of intense activity (also within our group).

\begin{figure}
\centering
\centerline{\includegraphics[width=\textwidth, height=6cm]{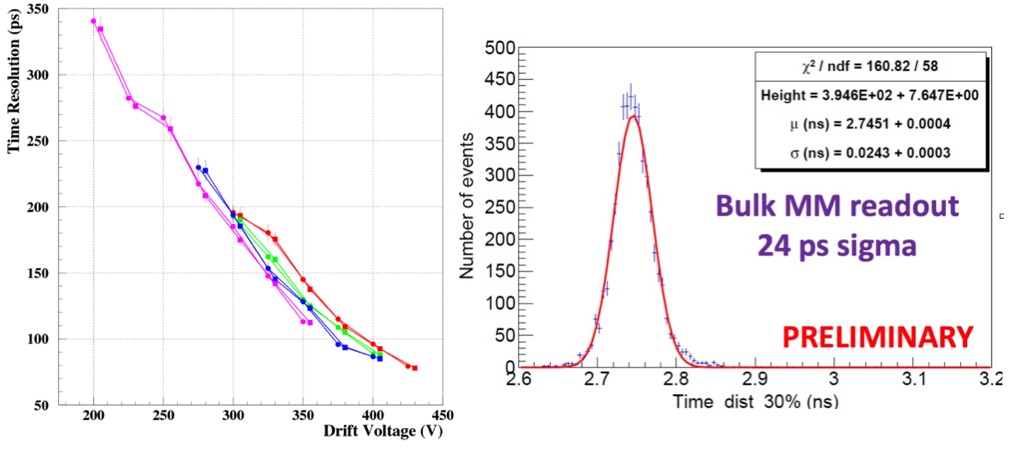}}
\caption{PICOSEC time resolution (single-photoelectron-Left and $\mu$-Right).}
\label{fig:BeamSchema}       
\end{figure}
\medskip 
 

\end{document}